\documentstyle[12pt]{article}
\begin{document}
\newcommand{\GeV}{\,{\rm GeV}\,}

\begin{titlepage}
\begin{center}
\null\vskip-1truecm
\begin{flushright}
IC/95/50 \\
hep-ph/9505273
\end{flushright}
\vskip1truecm
International Atomic Energy Agency\\
and\\
United Nations Educational Scientific and Cultural
Organization\\
\medskip
INTERNATIONAL CENTRE FOR THEORETICAL PHYSICS\\
\vskip2.3truecm
{\bf ON
THE FORM FACTORS OF THE $D^{\,+}_{\,s}\to \phi~\mu^{+}~\nu_{\mu}$
  DECAY}
\vskip1.2truecm
F. Hussain,\quad A.N. Ivanov$^1$,\quad N.I.
Troitskaya\footnote{\normalsize Permanent Address: Department of
 Theoretical
Physics, State Technical University, 195251 St. Petersburg, Russian
Federation.}\\
International Centre for Theoretical Physics, Trieste,
Italy.
\end{center}
\vskip1.5truecm
\centerline{\bf Abstract}
\bigskip

 We apply the infinite mass effective theory, when a heavy quark mass tends to
infinity, and Chiral perturbation theory at the quark level, based on the
extended Nambu -- Jona -- Lasinio model with linear realization of chiral
$U(3)\times U(3)$ symmetry, to calculate the form
factors of the $D^{\,+}_{\,s}\to\phi~\mu^+~\nu_{\,\mu}$ decay up to the
first order in current $s$ -- quark mass. The theoretical results
are compared with experimental data and found to be in good agreement.

\vfill
\begin{center}
MIRAMARE -- TRIESTE\\
May 1995
\end{center}
\vfill

\end{titlepage}

\newpage

\section*{\bf Introduction}

\hspace{0.2in} In our recent publications [1-3]  we considered the form
factors of the semileptonic  $\,D\,\to\,\bar{K}^{*}(\bar{K})\,e^+\,\nu_e\,$
decays both in the chiral limit [1,2] and at first order in current $s$-quark
mass expansion [3]. For the description of $D$-mesons we applied
the infinite mass effective theory (IMET) [4,5], when the $c$-quark mass
$M_c$ tends to infinity, i.e. $\,M_{\,c}\,\to\,\infty\,$. In the IMET, we
describe the long - distance physics within  Chiral perturbation theory at the
quark level (CHPT)$_q$ [6], based on the extended Nambu-Jona-Lasinio
(ENJL) model with linear realization of chiral \(U(3)\,\times\, U(3)\)
symmetry [7]. The IMET  supplemented by (CHPT)$_q$ has been successfully
applied to the description of the fine structure of the mass spectra
of non-strange $D\,(D^{\,\ast})$ and strange
$D^{\,+}_{\,s}\,(D^{\,\ast\,+}_{\,s})$ and leptonic constants, caused by
first order corrections in current-quark mass expansion [8]. The
computation of the probabilities of strong and electromagnetic $D^{\,\ast}$
decays performed within IMET and (CHPT)$_q$ has given good results
compared with experimental data [9].

In this paper we apply IMET and (CHPT)$_q$ to the calculation of the form
factors of the $D^{\,+}_{\,s}\to\phi~\mu^+~\nu_{\,\mu}$ decay, keeping
corrections up to first  order in current $s$-quark mass.

\section*{\bf 1. The form factors of the $D^{\,+}_{\,s}\to
\phi~\mu^{\,+}~\nu_{\,\mu}$ decay}

\hspace{0.2in} The amplitude of the
$D^{\,+}_{\,s}\to\phi~\mu^+~\nu_{\,\mu}$ decay is determined as follows

\begin{eqnarray}\label{label1}
M\, (D^{\,+}_{\,s}\to\phi~\mu^+~\nu_{\,\mu})\,=\,
-\frac{G_{\,F}}{\sqrt{2}}\,V^{\ast}_{\,c\,s}\,<\phi\,(Q)\vert\bar
s\,(0)\,\gamma_{\,\alpha}\,(1\,-\,\gamma^{\,5})\,c\,(0)\vert
D^{\,+}_{\,s}\,(p)>\,{\ell}^{\,\alpha}
\end{eqnarray}

\noindent where $\,G_{\,F} \,=\,1.166\times 10^{-5}\; {\rm GeV}^{-2}$ is the
Fermi weak coupling constant, $\vert V_{\,c\,s}\vert\,=\,0.975\,$ is the CKM
mixing matrix element, $s(0)$ and $c(0)$ are the $s$ and $c$ current
quark fields with $N$ colour degrees of freedom each, and
${\ell}^{\alpha}=\bar
u(k_{{\nu}_{\mu}})\gamma^{\alpha}(1-\gamma^{5})
v(k_{\mu^+})$ is the weak leptonic current.

The hadronic matrix element

\begin{eqnarray}\label{label2}
M_{\,\alpha}\,(D^{\,+}_{\,s}\,\to\,\phi)\,=\,<\phi\,(Q)\vert\bar
s\,(0)\,\gamma_{\,\alpha}\,(1\,-\,\gamma^{\,5})\,c\,(0)\vert
D^{\,+}_{\,s}\,(p)>
\end{eqnarray}

\noindent can be parametrized in terms of four form factors [3]

\parbox{11cm}{\begin{eqnarray*}
M_{\,\alpha}\,(D^{\,+}_{\,s}\,\to\,\phi)\,&=
&i\,a_{\,1}\,(q^{\,2})\,\epsilon^{\ast}
_{\,\alpha}\,(Q)\,-\,
i\,a_{\,2}\,(q^{\,2})\,(\epsilon^{\ast}\,(Q)\cdot p)\,(p\,+\,Q)_{\,\alpha}\,\\
&-&i\,a_{\,3}\,(q^{\,2})\,(\epsilon^{\ast}\,(Q)\cdot p)\,(p\,-\,Q)_{\,\alpha}\,
\\
&-&\,2\,b\,(q^{\,2})\,\varepsilon_{\,\alpha\,\beta\,\mu\,\nu}\,
\epsilon^{\,\ast\,\beta}\,(Q)\,p^{\,\mu}\,Q^{\,\nu}\,,\,
\end{eqnarray*}} \hfill
\parbox{1cm}{\begin{eqnarray}\label{label3}
\end{eqnarray}}

\noindent where $q^{\,2}$ is the square invariant mass of the lepton pair
such that $m^{\,2}_{\,\mu}\,\le q^{\,2}\,\le
(M_{\,D^{\,+}_{\,s}}\,-\,M_{\,\phi})^{\,2}$ where $M_{\,D^{\,+}_{\,s}}$,
$M_{\,\phi}$ and $m_{\,\mu}$ are the masses of the $D^{\,+}_{\,s}$
,$\,\phi$ and $\,\mu^{\,+}\,$ mesons respectively and $\epsilon^{\ast}$ is the
polarisation tensor of the outgoing $\phi$ meson.

We shall seek the form factors $a_{\,i}(q^{\,2})\;(i\,=\,1,2,3)$ and
$b\,(q^{\,2})$ in the form of an expansion in powers of the current
$s$-quark mass upto first order terms

\parbox{11cm}{\begin{eqnarray*}
a_{\,i}\,(q^{\,2}) &=& a^{(0)}_{\,i}\,(q^{\,2}) \,+\,
a^{(1)}_{\,i}\,(q^{\,2}) \,,\\
b\,(q^{\,2}) &=& b^{(0)}\,(q^{\,2}) \,+\,
b^{(1)}\,(q^{\,2}) \,
\end{eqnarray*}} \hfill
\parbox{1cm}{\begin{eqnarray}\label{label4}
\end{eqnarray}}

\noindent The form factors $a^{(0)}_{\,i}(q^{\,2})\;(i\,=\,1,2,3)$ and
$b^{(0)}\,(q^{\,2})$ are determined in the chiral limit $(ch.{\ell.})$.
They are calculated in the same way as the corresponding form factors
for the process
$D\to\bar{K}^{\,\ast}\,e^{\,+}\,\nu_{\,e}$ [1,3], that is

\parbox{11cm}{\begin{eqnarray*}
a^{(0)}_{\,1}\,(q^{\,2})&=&\sqrt{\frac{3}{8}}\;M_{\,\ast}\,\\
a^{(0)}_{\,2}\,(q^{\,2})&=&\sqrt{\frac{3}{8}}\;\frac{M_{\,\ast}}{M^{\,2}_{
\,D}}\,\biggl[\frac{q^{\,2}}{M^{\,2}_{\,D}\,-\,q^{\,2}}\,\\
&+&\frac{M^{\,2}_{\,D}\,-\,q^{\,2}}{M^{\,2}_{\,\ast}}
\left(\,1\,-\frac{2\,m\,M_{\,D}}{M^{\,2}_{\,D}\,-\,q^{\,2}}\right)
 {\ell
n}\left(\,1\,+\frac{M^{\,2}_{\,\ast}}{M^{\,2}_{\,D} \,-\,q^{\,2}}
\right)\biggr]\;,\\
a^{(0)}_{\,3}\,(q^2)&=
&-\,\sqrt{\frac{3}{8}}\;\frac{M_{\,\ast}}{M^{\,2}_{\,D}}\,
\biggl[\frac{2\,M^{\,2}_{\,D}\,-\,q^{\,2}}{M^{\,2}_{\,D}\,-\,q^{\,2}}\,\\
&-&\frac{M^{\,2}_{\,D}\,-\,q^{\,2}}{M^{\,2}_{\ast}}
\left(\,1\,-\frac{2\,m\,M_{\,D}}{M^{\,2}_{\,D}\,-\,q^{\,2}}\right)
 {\ell n}\left(\,1\,+\frac{M^{\,2}_{\,\ast}}{M^{\,2}_{\,D} \,-\,q^{\,2}}
\right)\biggr]\,,\\
b^{(0)}\,(q^{\,2})&=&\sqrt{\frac{3}{8}}\;\frac{1}{M_{\,\ast}}\,
{\ell n}\left(\,1\,+\frac{M^{\,2}_{\,\ast}}{M^{\,2}_{\,D} \,-\,q^{\,2}}
\right)\,.
\end{eqnarray*}} \hfill
\parbox{1cm}{\begin{eqnarray}\label{label5}
\end{eqnarray}}

\noindent Here we have denoted
$M_{\,\ast}\,=\,\sqrt{M_{\,D}\,\bar{v}^{\,\prime}/2}$ where
$M_{\,D}\,=\,1.86\;{\rm GeV}$ is the mass of the charmed pseudoscalar meson
at the chiral limit and $\bar{v}^{\,\prime}\,=\,4\,\Lambda \,=\,2.66
\;{\rm GeV}$. The parameter $\Lambda$ appears as the cut-off in the
Euclidian 3-dimensional momentum space evaluation
of constituent quark loop diagrams. This cut-off $\Lambda$ is connected with
the scale of spontaneous breaking of chiral symmetry (SBCS) in (CHPT)$_q$ via
the relationship $\Lambda\,=\,\Lambda_{\,\chi}/\sqrt{2}\,=\,0.67\;{\rm
GeV}$ at $\Lambda_{\,\chi}\,=\,0.94\;{\rm GeV}$ [6].

The form factors $a^{(1)}_{\,i}(q^{\,2})\;(i\,=\,1,2,3)$ and
$b^{(1)}\,(q^{\,2})$ are determined by the matrix element [3]

\parbox{11cm}{\begin{eqnarray*}
&&M^{(1)}_{\,\alpha}\,(D^{\,+}_{\,s}\,\to\,\phi)\,= \,-\,i\,m_{\,0\,s}
 \\
&& \times \int\,d^{\,4}\,x\,<h\,(Q)\vert {\rm T}(\,{\bar
s}\,(x)\,s\,(x)\,\bar s\,(0)\,\gamma_{\,\alpha}\, (1\,-\,\gamma^{\,5})
 \,
c\,(0)) \vert D^{\,+}_{\,s}\,(p)>_{ch.{\ell.}}\,.
\end{eqnarray*}} \hfill
\parbox{1cm}{\begin{eqnarray}\label{label6}
\end{eqnarray}}

\noindent In accordance with the procedure expounded in [1-3,8,9] we can
reduce the matrix element (\ref{label6}) to the expression

\begin{eqnarray}\label{label7}
&&M^{(1)}_{\,\alpha}\,(D^{\,+}_{\,s}\,\to\,\phi)\,=\,g_{\,D}\,m_{\,0\,s}\,
i\
,\int\,d^{\,4}\,x\,\int^{\infty}_{-\infty}\,d\,z_{\,0}\,\theta\,(\,-\,z_{\,0
}\,)\times\nonumber\\
&&\times<\,\phi\,(Q)\vert {\rm T}
(\,\bar{s}\,(x)\,s\,(x)\,\bar{s}(0)\,\gamma_{\,\alpha}\,
\left(\frac{1\,+\,\gamma^{\,0}}{2}\right)\,
\gamma^{\,5}\,s\,(z_{\,0},\vec{0}\,))\vert{0}>_{ch.{\ell.}}
\end{eqnarray}

\noindent obtained at leading order in the large $N$ and $M_{\,c}$ expansion.
The coupling constant $\,g_{\,D}\,$ has been calculated in [9]

\begin{equation}\label{label8}
g_{\,D}\,=\,\frac{2\,\sqrt{2}\,\pi}{\sqrt{N}}\,\left(\frac{M^{\,2}_{\,D}}{
M_
c\,\bar v^{\,\prime}}\right)^{1/2}
\end{equation}

\noindent The r.h.s. of (\ref{label7}) involves only the light quark
fields. Therefore for the evaluation of (\ref{label7}) we can apply
(CHPT)$_q$ [1-3,6,8,9]. Since the leading order of the r.h.s. of
(\ref{label7}) in current quark mass expansion is fixed by the factor
\(\,m_{\,0\,s}\,\), so the matrix element \(\,<\,\phi\,(Q)\vert\,{\rm
T}(\ldots)\,\vert{0}>\,\) has to be calculated in the chiral limit.

In order to evaluate the matrix element (\ref{label7}) let us compare this
with the matrix element
$\,M^{(1)}_{\,\alpha}\,(D\,\to\,\bar{K}^{\,\ast})\,$ describing the
$D\,\to\,\bar{K}^{\,\ast}$ transition at the first order in current $s$-
quark mass expansion [3]

\begin{eqnarray}\label{label9}
&&M^{(1)}_{\,\alpha}\,(D\,\to\,\bar{K}^{\,\ast})\,=\,g_{\,D}\,m_{\,0\,s}\,
i\
,\int\,d^{\,4}\,x\,\int^{\infty}_{-\infty}\,d\,z_{\,0}\,\theta\,(\,-\,z_{\,0
}\,)\times\nonumber\\
&&\times<\,\bar{K}^{\,\ast}\,(Q)\vert {\rm T}
(\,\bar{s}\,(x)\,s\,(x)\,\bar{s}(0)\,\gamma_{\,\alpha}\,
\left(\frac{1\,+\,\gamma^{\,0}}{2}\right)\,\gamma^{\,5}\,q\,
(z_{\,0},\vec{0}\,))\vert{0}>_{\rm ch.l.}
\end{eqnarray}

\noindent where $q\,=\,u$ or $d$ for $D^{\,0}$ or $D^{\,+}$, respectively.

By applying the formulas of quark conversion (Ivanov [6]) one can show that,
between matrix elements $M^{(1)}_{\,\alpha}\,(D^{\,+}_{\,s}\,\to\,\phi)$
and $\,M^{(1)}_{\,\mu}\,(D\,\to\,\bar{K}^{\,\ast})\,$,  there is the
relationship

\begin{eqnarray}\label{label10}
M^{(1)}_{\,\alpha}\,(D^{\,+}_{\,s}\,\to\,\phi)
\,=\,2\,M^{(1)}_{\,\alpha}\,(D\,\to\,\bar{K}^{\,\ast})\,.
\end{eqnarray}

\noindent Readers can verify this relationship by noting
that the $\phi$ meson possesses the quark structure $(\bar{s}\,s)$.

By virtue of the relationship (\ref{label10}) the form factors
$a^{(1)}_{\,i}(q^{\,2})\;(i\,=\,1,2,3)$ and $b^{(1)}\,(q^{\,2})$
 read [3]

\parbox{11cm}{\begin{eqnarray*}
a^{(1)}_{\,1}\,(q^{\,2})&=&{\sqrt{3}}\;\frac{m_{\,0\,s}}{M_{\ast}}\,
\frac{\bar v}{4\,m}\,M_{\,D}\;
{\ell n}\left(\frac{{\bar v}^{\,\prime}}{4\,m}\right)\,\\
a^{(1)}_{\,2}\,(q^{\,2})&=&-\,a^{(1)}_{\,3}\,(q^{\,2})\,=
\,b^{(1)}\,(q^{\,2})\,\\
b^{(1)}\,(q^{\,2})&=&\sqrt{3}\;\frac{m_{\,0\,s}}{M_{\ast}}\,\frac{\bar
v}{4\,m}\,\frac{M_{\,D}}{M^{\,2}_{\,D}\,-\,q^{\,2}}\,\Big[\,1\,-\,{\ell
n}\left(1\,+\,\frac{M^{\,2}_{\ast}}{M^{\,2}_{\,D}\,- \,q^{\,2}}
\right)\Big]\,.
\end{eqnarray*}} \hfill
\parbox{1cm}{\begin{eqnarray}\label{label11}
\end{eqnarray}}

\noindent Now we can get the numerical values of the form factors
 at $\,q^{\,2}\,=\,0\,$

\parbox{11cm}{\begin{eqnarray*}
\begin{array}{llcl}
a_{\,1}\,(0)&=&a^{(0)}_{\,1}\,(0)\,+\,a^{(1)}_{\,1}\,(0)\,=~\,0.96\,
+\,0.28\,=~\,1.24\;({\rm GeV})\,,&  \\
a_{\,2}\,(0)&=&a^{(0)}_{\,2}\,(0)\,+\,a^{(1)}_{\,2}\,(0)\,=~\,0.14\,
+\,0.06\,=~\,0.20\;({\rm GeV})^{-\,1}\,,& \\
a_{\,3}\,(0)&=&a^{(0)}_{\,3}\,(0)\,+\,a^{(1)}_{\,3}\,(0)\,=\,-\,0.42\,-\,
0.06\,=\,-\,0.48\;({\rm GeV})^{-\,1}\,,& \\
b\,(0)&=&b^{(0)}\,(0)\,+\,b^{(1)}\,(0)\,=~\,0.21\,
+\,0.06\,=~\,0.27\;({\rm GeV})^{-\,1}\,. &
\end{array}
\end{eqnarray*}} \hfill
\parbox{1cm}{\begin{eqnarray}\label{label12}
\end{eqnarray}}

\noindent One finds that the first order corrections in current $s$-
quark  mass expansion are between $14$ and $43 \%$. The form factors
$\,a_{\,i}\,(q^{\,2})\;(\,i\,=\,1\,,\,2\,, \,3)\,$ and $\,b\,(q^{\,2})\,$
are connected with the standard form factors $\,A_{\,i}\,(q^{\,2})=
 \;
(\,i\,=\,1\,,\,2\,,\,3)\,$ and $\,V\,(q^{\,2})\,$ via the relations [1]

\parbox{11cm}{\begin{eqnarray*}
A_{\,1}\,(0)&=&\frac{1}{M_{\,D_{\,s}}\,+\,M_{\,\phi}}a_{\,1}\,(0)\,=\,~0.43\,\\
A_{\,2}\,(0)&=&(M_{\,D_{\,s}}\,+\,M_{\,\phi})\;a_{\,2}\,(0)\,=\,~0.60\,\\
A_{\,3}\,(0)&=&(M_{\,D_{\,s}}\,+\,M_{\,\phi})\;a_{\,3}\,(0)\,=\,-\,1.44\,\\
V\,(0)&=&(M_{\,D_{\,s}}\,+\,M_{\,\phi})\;b\,(0)\,=\,~0.81\,
\end{eqnarray*}} \hfill
\parbox{1cm}{\begin{eqnarray}\label{label13}
\end{eqnarray}}

\noindent where $M_{\,D_{\,s}}\,=\,1.97\;{\rm GeV}$ and
$\,M_{\,\phi}\,=\,1.02\;{\rm GeV}\,$ [10].

The theoretical values are in good agreement with the experimental data
 [11,12]

\parbox{11cm}{\begin{eqnarray*}
 (R_{\,v})_{\,\rm th}\,=\,\frac{V(0)_{\,\rm th}}{{A_{\,1}}(0)_{\,\rm
th}}\,=\,1.9,&&\quad(R_{\,v})_{\,\rm
exp}\,=\,
\left\{\begin{array}{r}1.8\,\pm\,0.9\,\pm\,0.1\;[11] \\ 1.4\,\pm\,0.5\,\pm\,0.3
\;[12]\end{array} \right.\\
(R_{\,2})_{\,\rm th}\,=\,\frac{{A_{\,2}}(0)_{\,\rm
th}}{{A_{\,1}}(0)_{\,\rm th}}\,=\,1.4,&&\quad(R_{\,2})_{\,\rm
exp}\,=\,\left\{\begin{array}{r}
1.1\,\pm\,0.6\,\pm\,0.1\;[11] \\ 0.9\,\pm\,0.6\,\pm\,0.3 \;[12]\end{array}
\right.
\end{eqnarray*}} \hfill
\parbox{1cm}{\begin{eqnarray}\label{label14}
\end{eqnarray}}

\section*{\bf Conclusion}

\hspace{0.2in}  By applying IMET
supplemented by (CHPT)$_q$ we have evaluated the form factors of the
$D^{\,+}_{\,s}\,\to\,\phi\,\mu^{\,+}\,\nu_{\,\mu}$ decay upto first
order in current $s$-quark mass. The theoretical predictions
compare reasonably well with experimental data. The proposed approach allows
us to set up a relationship (\ref{label10}) between chiral corrections to the
form factors of the decays $D^{\,+}_{\,s}\,\to\,\phi\,\mu^{\,+}\,\nu_{\,\mu}$
and $D\,\to\,\bar{K}^{\,\ast}\,e^{\,+}\,\nu_{\,e}$. Unfortunately, the
possibility of the experimental investigation of the relationship
(\ref{label10}) goes beyond the available accuracy of present day experimental
abilities.

\section*{\bf Acknowledgements}

\hspace{0.2in} ANI and NIT would like to thank Professor Abdus Salam, the
International Atomic Energy Agency and UNESCO for hospitality at the
International Centre for Theoretical Physics, Trieste. We acknowledge with
pleasure fruitful discussions with Prof. G. E. Rutkovsky.

\newpage

\begin{center}
\section*{\bf References}
\end{center}
\vspace{0.5in}
\begin{description}

\item{[1]}~F. Hussain, A. N. Ivanov and N. I. Troitskaya, Phys. Lett.
 {\bf
B 329} (1994) 98; ibid. {\bf B 334} (1994) E450.
\item{[2]}~A. N. Ivanov, N. I. Troitskaya and M. Nagy, Phys. Lett.
 {\bf B
339} (1994) 167.
\item{[3]}~F. Hussain, A. N. Ivanov and N. I. Troitskaya, Phys. Lett.
{\bf B 348} (1995) 609.
\item{[4]}~E. Eichten and F. L. Feinberg, Phys. Rev. {\bf D 23} (1981)
 2724;
\item{~~~}~E. Eichten, Nucl. Phys. {\bf B 4} (Proc.Suppl.) (1988)
 70;
\item{~~~}~M. B. Voloshin and M. A. Shifman, Sov. J. Nucl. Phys.
 {\bf 45}
(1987) 292;
\item{[5]}~H. D. Politzer and M. Wise, Phys. Lett. {\bf B 206} (1988)
 681;
ibid. {\bf B 208} (1988) 504.
\item{[6]}~A. N. Ivanov, M. Nagy and N. I. Troitskaya,
Int. J. Mod. Phys. {\bf A 7} (1992) 7305;
\item{~~~}~A. N. Ivanov, Int. J. Mod. Phys. {\bf A 8} (1993) 853;
\item{~~~}~A. N. Ivanov, N. I. Troitskaya and M. Nagy,
Int. J. Mod. Phys.{\bf A 8} (1993) 2027; 3425;
\item{~~~}~A. N. Ivanov, N. I. Troitskaya and M. Nagy, Phys. Lett.
 {\bf B
308} (1993) 111;
\item{~~~}~A. N. Ivanov and N. I. Troitskaya, ``$\pi$- and $a_1$-
 meson
physics in current algebra at the quark level", ICTP, Trieste, preprint
IC/94/10, January 1994 (to appear in Nuovo Cimento A).
\item{[7]}~Y. Nambu and G. Jona-Lasinio, Phys. Rev. {\bf 122}
 (1961)
345; ibid. {\bf 124} (1961) 246.
\item{~~~}~T. Eguchi, Phys. Rev. {\bf D 14} (1976) 2755;
\item{~~~}~K. Kikkawa, Progr. Theor. Phys. {\bf 56} (1976) 947;
\item{~~~}~H. Kleinert, Proc. of Int. Summer School of Subnuclear
Physics, Erice 1976, Ed. A. Zichichi, p.289.
\item{[8]}~A. N. Ivanov and N. I. Troitskaya, Phys. Lett. {\bf B 342}
(1995) 323.
\item{[9]}~A. N. Ivanov and N. I. Troitskaya, Phys. Lett. {\bf B 345} 175.
\item{[10]}~Particle Data Group, Phys. Rev. {\bf D 50} (1992) No.3, Part I.
\item{[11]}~P. L. Fabertti et. al., Phys. Lett. {\bf B 328} (1994) 187.
\item{[12]}~P. Avery et. al., CLEO Collaboration, Phys. Lett. {\bf B 337}
(1994) 405.
\end{description}

\end{document}